\documentstyle[preprint,aps]{revtex}
\begin{document}

\draft

\preprint{$
\begin{array}{l}
\mbox{UCD--95--22}\\[-3mm]
\mbox{July~1995} \\   
\end{array}
$}

\title{Triple Electroweak Gauge-Boson Production \\
       at  Fermilab Tevatron Energies}

\author{Tao Han\footnote{e-mail: than@ucdhep.ucdavis.edu}
 and Ron Sobey\footnote{e-mail: rsobey@ucdhep.ucdavis.edu} }

\address{Davis Institute for High Energy Physics, \\
         Department of Physics,
         University of California, Davis, CA 95616, USA}
\maketitle
\begin{abstract}

We calculate  the  three gauge-boson production in the Standard Model
at Fermilab Tevatron energies.  At $\sqrt s=2$ TeV in $p\bar p$ collisions,
the cross sections for the triple gauge-boson production
are typically of  order 10 femtobarns (fb).
For  the pure leptonic  final states from the gauge-boson decays
and with some minimal cuts on final state photons,
the cross sections for
$p \bar p \rightarrow W^\pm \gamma\gamma, Z\gamma\gamma$
and $W^+W^- \gamma$ processes are of order a few  fb,
resulting in a few dozen clean leptonic events
for an integrated luminosity of 10 fb$^{-1}$.
The pure leptonic modes from other  gauge-boson channels
give significantly smaller rate. Especially, the trilepton modes
from $W^+W^-W^\pm$ and  $t \bar t W^\pm \rightarrow W^+W^-W^\pm$
yield a cross section  of   order  0.1 fb  if there is no
significant Higgs boson contribution. For a Higgs boson with
$m_H^{} \simeq 2M_Z^{}$,
the triple massive-gauge-boson  production rate
could be enhanced by a factor of  $4-6$.

\end{abstract}


\narrowtext

\section{Introduction}

Ever since the experimental discovery of the electroweak vector bosons
$W^\pm$ and $Z$ (generically denoted by $V$ henceforth, unless
specified otherwise)
at the CERN $\bar p p$ collider, studies on their production
in collider experiments have provided important means to examine
the electroweak Standard Model (SM), and to explore new physics
beyond the SM  \cite{overview}.
The production of  two gauge-bosons \cite{pair}  at
hadron and $e^+e^-$ colliders  may provide a crucial test
of the non-Abelian three-gauge-boson interactions \cite{dpfww},
because of  intimate cancellation  in the  SM
between the contributions of
gauge-boson exchanges in the $s$-channel and fermion exchanges
in the $t$- and $u$-channels.
The production of three  gauge-bosons \cite{triple1,triple2,tripleee}
involves four-gauge-boson
interactions and should  be sensitive to these
quartic vertices \cite{quartic}.  Recent  experimental
studies of  gauge-boson pair events at hadron colliders
have set limits on  anomalous couplings
for the three-gauge-boson interactions
from the direct production channels \cite{cdfd0}.
Further improvements on those studies are anticipated from
future experiments with the Tevatron Main Injector
(an annual integrated luminosity of 1 fb$^{-1}$).
A possible further upgrade of the Tevatron
(a luminosity  of  order 10$^{33}$cm$^{-2}$s$^{-1}$ \cite{ditev}
or about 10  fb$^{-1}$/yr, and a possible
energy upgrade to 4 TeV  \cite{upgrade}),  especially
the LHC (a luminosity of $10-100$ fb$^{-1}$ \cite{atlascms})
may also add to these studies \cite{hadron}.
Experiments at LEP II will study the process
$e^+e^- \rightarrow W^+W^-$ in great detail \cite{lepii}.
The future  $e^+e^-$ Linear Colliders will explore not only the
processes of pair gauge-boson production, but also the triple
gauge-boson production \cite{NLCWW,quartic}.
Furthermore, since heavy particles often decay to gauge-bosons
as a final state, it is important to examine the multi-gauge-boson
production in searching for  new physics. The most  notable
example is that  the  recent  studies on
$WW(\rightarrow l\nu,l\nu)$+2-jet and $W(\rightarrow l\nu)$+4-jet events
at the Fermilab Tevatron
have resulted in the discovery of the elusive top quark \cite{top}.
Other examples include a heavy Higgs boson decay
$H^0  \rightarrow W^+W^-,ZZ$ \cite{atlascms,higgstev};
heavy chargino and neutralino decays and squark/gluino cascade decays
in supersymmetric theories \cite{susy};
longitudinal gauge-boson strong scattering \cite{sews};
new heavy gauge-boson decays \cite{tomr};
and other exotic heavy particles that may decay to electroweak
gauge-bosons \cite{exotica}.

Motivated by the great success of the Fermilab Tevatron experiments,
we carry out  the calculation for three
electroweak-gauge-boson  production in the SM for
the energy range  $1.8-4$ TeV,  relevant to future experiments
at   a  Tevatron upgrade for luminosity and center
of mass energy \cite{ditev,upgrade}.
We have considered all combinations for the three
electroweak-gauge-boson processes (except for the pure QED process
of three-photon production):
\begin{eqnarray}
\label{EQ:VGG}
p \bar p &\rightarrow & W^\pm \gamma \gamma, \ Z \gamma \gamma, \\
\label{EQ:VVG}
p \bar p  &\rightarrow & W^+W^-\gamma, \ W^\pm Z \gamma, \ ZZ\gamma, \\
\label{EQ:VVV}
p \bar p  &\rightarrow & W^+W^- W^\pm, \ W^+W^- Z,  \ W^\pm ZZ, \  ZZZ
\end{eqnarray}
Since a top quark almost exclusively decays to a $W^+b$ final state,
there will be significant contribution to $W^+W^-$ production from
$t\bar t$ decays. We therefore also include the top-quark channels
\begin{eqnarray}
p \bar p &\rightarrow & \ t \bar t \gamma, \ t \bar t  W^\pm,
\  t \bar t  Z \quad {\rm with}   \quad \ t \bar t \rightarrow W^+W^- .
\label{EQ:TTV}
\end{eqnarray}
These processes of Eqs.~(\ref{EQ:VGG}--\ref{EQ:TTV})
have rather small production cross sections at Tevatron energies,
typically being of order 10 fb, about 100 times
smaller than the gauge-boson pair production. However, as
we will see, it is  possible to observe those processes
at  an upgraded Tevatron.  They must be quantified  in order to
explore the quartic gauge-boson self-interactions. Moreover, they
need to be reliably estimated in searching for new physics,
such as in the multi-charged lepton channels in SUSY
searches \cite{susy}.

In Sec.~II, we present  the cross sections  of  the
processes in Eqs.~(\ref{EQ:VGG}--\ref{EQ:TTV}).
In Sec.~III,  we  discuss the results with leptonic final states
from the gauge-boson decays and some typical kinematical distributions.
We finally summarize our results in Sec.~IV.

\section{Cross Sections for Triple Gauge-Boson Production
at  Tevatron Energies}

The processes listed in Eqs.~(\ref{EQ:VGG}--\ref{EQ:TTV})
have been individually studied in the literature,
but only  in $pp$ collisions at Supercollider energies.  The
processes $W^\pm \gamma \gamma$ \cite{wgmgm}
and $Z\gamma \gamma$ \cite{zgmgm}
in Eq.~(\ref{EQ:VGG}) have been calculated
as backgrounds to the intermediate Higgs boson signal
for $H^0 \rightarrow \gamma\gamma$.
These in  Eq.~(\ref{EQ:VVG}) were first evaluated in Ref. \cite{triple2}
and those of  Eq.~(\ref{EQ:VVV})  in Refs. \cite{triple1,triple2}.
Calculations for the $t\bar t W^\pm$, $t\bar t Z$ and $t\bar t \gamma$
processes first  appeared in Refs. \cite{wtt}, \cite{ztt} and \cite{gtt},
respectively.  We have been able to reproduce the total cross sections
in the literature whenever available to compare.

In this section, we present the total cross sections for all the
production  processes in Eqs.~(\ref{EQ:VGG}--\ref{EQ:TTV})
in $p \bar p$ collisions at Tevatron energies.
In our numerical calculations,
we have adopted the helicity amplitude techniques, following
the schemes in Refs. \cite{hz} and \cite{alan}.
We have chosen the following input mass parameters \cite{databook}:
$M_W=80.22$ GeV, $M_Z=91.187$ GeV,  $m_t=175$ GeV,
and $\alpha_{em}=1/128$.
We use the parton distribution functions Set A of
Martin-Roberts-Stirling \cite{mrsa}. For the processes
of Eqs.~(\ref{EQ:VGG}--\ref{EQ:VVV}),  the factorization
scale is set at the parton c.~m.~energy $\sqrt {\hat s}$; while
for the processes of Eq.~(\ref{EQ:TTV}), we set the renormalization
scale in $\alpha_s$ and the  factorization scale both at twice
of the top-quark mass $2m_t$.

Figure~\ref{FIG:VGG} shows the total cross sections for
processes with one or two direct photons,
Eqs.~(\ref{EQ:VGG}--\ref{EQ:VVG}).  To avoid the infrared divergence
for the final state photons in our tree-level calculation
and to roughly simulate the experimental detector coverage,
we have imposed some minimal cutoff on the
transverse momentum ($p_T^\gamma$) and the
pseudorapidity ($\eta^\gamma$) as
\begin{eqnarray}
p_T^\gamma > 10  \  {\rm  GeV} ;
\quad {\rm and}   \quad |\eta^\gamma| < 2.5 ,
\label{EQ:GCUT}
\end{eqnarray}
where $\eta^\gamma=\ln \cot (\theta^\gamma/2)$, with $\theta^\gamma$
the polar angle of the photon in the parton c.~m.~frame with respect to
the proton direction.
No cuts are applied to the  $W^\pm$ and $Z$ throughout this paper;
but we will comment in Sec.~II on the effects if experimental acceptance cuts
are imposed.
We see that at  $\sqrt s$=2 TeV,
the total cross sections are about 7 fb for $ZZ\gamma$ (lowest)
and 44 fb for $W^+W^-\gamma$ and  $Z\gamma\gamma$ (highest).
For  comparison, the cross sections  for  $Z \gamma$ and $W^\pm \gamma$
production at  2 TeV are about 16 pb and 17 pb, respectively.
Increasing the c.~m.~energy from 2 TeV to 4 TeV
would enhance the production cross sections
for the triple gauge-boson production by about a factor of 3.
Naively, one would expect that
the $W^\pm$-related cross sections should be larger than those of $Z$,
like in the case of Drell-Yan production of $W^\pm$  and $Z$ bosons,
due to the weak charge-current coupling being stronger
than the neutral-current coupling.
It is therefore surprising to see that $W^\pm \gamma \gamma$
rate is even smaller than that of  $Z \gamma \gamma$. In fact, this is
due to  some severe gauge cancellation in $W^\pm\gamma\gamma$
process, a phenomenon called radiation amplitude zero \cite{raz}.
The fact that the $W^\pm Z\gamma$ cross section
tends to be somewhat smaller than naively expected
(compared to  $W^+ W^- \gamma$  and $Z Z\gamma$ cases)
may be also traced to a radiation
amplitude zero for $W^\pm \gamma$  process \cite{raz} and an
approximate zero in  $W^\pm Z$  process \cite{wzzero}.

Figure~\ref{FIG:VVV} presents  the total cross sections for
processes with three massive bosons, Eq.~(\ref{EQ:VVV}).
The cross sections are generally smaller by about an
order of magnitude if we replace a photon [with our
choice of the cuts  in Eq.~(\ref{EQ:GCUT})]  by a $Z$.
The cross section rates at  $\sqrt s$=2 TeV are about 0.5 fb
for $ZZZ$ process (lowest)
and 6 fb for $W^+W^-W^\pm$ process (highest).
Again for comparison, the cross sections
for  $ZZ$, $W^\pm Z$ and $W^+ W^-$
production at  2 TeV are about 1.1 pb,  2.5 pb and 9.2 pb,
respectively. It is interesting to note that the Higgs boson
in the SM  contribute to the processes of  Eq.~(\ref{EQ:VVV})
through $H^0 \rightarrow VV$.
In the calculation for Fig.~\ref{FIG:VVV},
we have taken the Higgs boson mass
$m_H^{}=100$ GeV, for which  there is no significant contribution
from $H^0$ in $VVV$ production.  If  the Higgs boson mass is just above
the $VV$ pair threshold, the enhancement to $VVV$ production
could be sizeable.  The  Higgs boson effects are
demonstrated  in Fig.~\ref{FIG:HIGGS},
where the total cross sections are plotted
versus $m_H^{}$ for $\sqrt s=2$ TeV.
We see that the increase right above the threshold
is as much as a factor of 4 for $W^\pm ZZ$ and  $ZZZ$ channels through
$H^0 \rightarrow ZZ$,  and a factor of 6
for $W^+W^-W^\pm$ and $W^+W^-Z$ through $H^0 \rightarrow W^+W^-$.

For the top-quark induced triple gauge-boson production
(via $t\bar t \rightarrow W^+W^-$ with a 100\% branching fraction),
the $t \bar t W^\pm$ production goes only through $q\bar q'$-annihilation
diagrams, in which one of the light quarks radiates  a  $W$;
while  $t \bar t Z$ and $t \bar t  \gamma$ get contributions
from both $q\bar q$-annihilation and $gg$-fusion.
At Tevatron energies, the valence quark contributions dominate
and the $gg$-fusion only contributes a few percent.
Cross sections for those channels are presented in Fig.~\ref{FIG:TTV}.
Formally, these processes are of order $\alpha_s^2 \alpha$, compared
to the electroweak triple gauge-boson cross section of order $\alpha^3$.
However, due to phase space suppression from the large top-quark mass,
the production cross sections in Fig.~\ref{FIG:TTV} are comparable to
that of $W^+W^-\gamma$ in Fig.~\ref{FIG:VGG} and
those of $W^+W^- W^\pm$ and $W^+W^- Z$ in Fig.~\ref{FIG:VVV},
although the rates for the $t \bar t V$ processes increase more rapidly
than those of $VVV$ for higher energies. Since the heavy top quark
often gives a hard $b$-jet  as a decay product, one may be able to
separate the electroweak $W^+W^- V$ process from
$t \bar t  V\rightarrow W^+W^- V $ by examining the additional
jet activities, if desired.
We have not included the processes
resulting from $H^0 \rightarrow t \bar t$, such as
\begin{eqnarray}
\label{EQ:VH}
p \bar p  \rightarrow   W^\pm H^0 \rightarrow W^\pm t \bar t
\quad  {\rm and} \quad
p\bar p \rightarrow  Z H^0 \rightarrow Z t \bar t .
\end{eqnarray}
The decay branching fraction of
$H^0 \rightarrow t \bar t$ is never larger than those of
$H^0 \rightarrow  VV$  which we included in the calculations
for Eq.~(\ref{EQ:VVV}). Also, the cross sections of Eq.~(\ref{EQ:VH})
would be large only if $m_H^{} \geq 2m_t \simeq 400$ GeV, where the
total rate at Tevatron energies must be very small due to the phase
space suppression.

\section{Discussions}

The predicted cross sections
for triple gauge-boson production in the SM, as shown in
Figs.~\ref{FIG:VGG}--\ref{FIG:TTV}, at Tevatron energies
may result in  a sizeable number of  events  if  the luminosity is
sufficiently high.
With  the Tevatron Main Injector, an annual integrated luminosity
of order 1 fb$^{-1}$ is envisioned.  A further upgrade of the luminosity
to reach 10 fb$^{-1}/$year may become reality \cite{ditev,upgrade}.
However, in the hadron collider environment,
the large QCD backgrounds may render the observation impossible
for those events if  one or more
of the gauge bosons decay hadronically. Individual channels with
hadronic decays should be studied on a case by case basis.
For simplicity, we only discuss the cleanest channels with pure
leptonic decays of the gauge bosons.
Taking the relevant branching fractions to be  \cite{databook}
%
%
\begin{eqnarray}
\label{EQ:BR}
BR(W^+ \rightarrow l^+ \nu)=21\%, \quad
BR(Z\rightarrow l^+ l^-)=6.7\%,
\end{eqnarray}
where $l=e$ and $\mu$, we have
\begin{eqnarray}
\nonumber
&BR&(W^+W^-\rightarrow l^+ \nu l^- \bar\nu)=4.6\%, \quad
BR(ZZ\rightarrow l^+ l^- l^+ l^-)=0.45\%, \\
&BR&(W^+ Z\rightarrow l^+ \nu l^+ l^-)=
BR(W^- Z\rightarrow l^- \nu l^+ l^-)=1.4\%,
\label{EQ:BRVV}
\end{eqnarray}
and
\begin{eqnarray}
\nonumber
&BR&(W^+W^-W^+\rightarrow l^+ \nu l^- \bar\nu l^+ \nu)=0.98\%, \quad
BR(W^+W^-Z\rightarrow l^+ \nu l^- \bar\nu l^+ l^-)=0.31\%, \\
&BR&(W^+ ZZ\rightarrow l^+ \nu l^+ l^- l^+ l^-)= 9.7 \times10^{-4}, \quad
BR(ZZ Z\rightarrow  l^+ l^- l^+ l^-  l^+ l^-)=3.0 \times 10^{-4}.
\label{EQ:BRVVV}
\end{eqnarray}
For the direct photons, we assume that they
can  be identified effectively as isolated showers
in the electromagnetic calorimeter.

With these branching fractions, one can easily estimate the cross
section rate for a given leptonic mode based on the total cross
sections presented in Figs.~\ref{FIG:VGG}--\ref{FIG:TTV}.
We summarize the cross section rates in units of fb
for  all channels  in two Tables for $\sqrt s$=2 TeV.
Table~\ref{T:ONE} presents the processes with direct
photons and Table~\ref{T:TWO} only those with massive $V$'s.
Entries in the first row list the total cross sections
read off directly from the figures; those in the second row give the
cross section rates with the relevant  leptonic  branching fractions.
The letter $X$ there denotes the missing neutrinos.
We see that  with an integrated luminosity of 10 fb$^{-1}$, one
would expect a few dozen of pure leptonic events for
$W^\pm \gamma\gamma, Z\gamma\gamma$
and $W^+W^- \gamma$, making the studies for anomalous quartic
couplings at  the upgraded Tevatron possible.
The other channels are of very small event rate, even though
a Higgs boson with $m_H^{} \geq 2M_V$ could enhance the
$VVV$ production rates by a factor 4 -- 6.
The trilepton channels of
$W^+ W^- W^\pm$ and  $t \bar t  W^\pm \rightarrow l^+ l^- l^\pm X$
are of special interest since  they are
the backgrounds to the benchmark SUSY
processes for  gaugino signals.
The total rate of about 0.1 fb shown in Table~\ref{T:TWO}
would  not  seem  to pose a severe problem for the trilepton
SUSY signals  \cite{susy}.

We have thus far  simply imposed a minimal $\eta^\gamma$-cutoff ($<2.5$)
on the final state photons.
It  is informative  to examine the photon pseudorapidity distribution.
Figure~\ref{FIG:ETA}(a)  shows the $|\eta^\gamma|$
distribution for $W^+W^- \gamma, Z\gamma \gamma$
and $W^\pm \gamma \gamma$ processes,
and Fig.~\ref{FIG:ETA}(b) presents the $|\eta^\gamma_{\rm max}|$
(the larger one of the two $|\eta^\gamma|$) distribution
for $W^\pm \gamma \gamma$ and $Z\gamma \gamma$
processes, with a cut $p_T^\gamma > 10$ GeV.
We see that if we tighten up the $\eta^\gamma$-cut
to be 1, the total rates would be reduced to
22 fb  (by 49\%) , 13 fb  (73\%) and 5.4  fb (75\%)
for  $W^+W^- \gamma, Z\gamma \gamma$
and $W^\pm \gamma \gamma$ processes, respectively (cf. Table~\ref{T:ONE}).

It is important to note that in Fig.~\ref{FIG:ETA}(a)
the cross section rate for
$W^\pm \gamma \gamma$ in the central scattering region (small
pseudorapidity) is much  lower  than that for
$Z \gamma \gamma$ and $W^+ W^-\gamma$: a direct reflection
of  the radiation amplitude  zeros in the $W^\pm \gamma\gamma$ processes.
When the two photons are collinear,
there would be an exact radiation amplitude zero \cite{raz}  in the
parton c.~m.~frame.
For the dominant process $u_1 \bar d_2 \rightarrow W^+ \gamma\gamma$,
it is located at
\begin{eqnarray}
\label{EQ:ZERO}
\cos \theta^\gamma= -\frac {Q_u + Q_d} {Q_u-Q_d},
\end{eqnarray}
which is at $ -1/3$;  while the zero would be
located at  $ +1/3$ for  $d_1 \bar u_2 \rightarrow W^- \gamma\gamma$.
It is these zeros that make the $W^\pm \gamma\gamma$ cross section small
in the central region. More importantly, the zeros are due to subtle gauge
cancellation in the SM and thus sensitive to any anomalous couplings
beyond the SM. We will examine this aspect in more detail in another
publication.

Similar to Fig.~\ref{FIG:ETA},  Figure~\ref{FIG:PT}  presents the
transverse momentum distribution
of the photon(s) for these three processes, with a cut
$|\eta^\gamma|<2.5$.  As expected, the distributions fall off sharply.

Finally, we present the maximum  rapidity distribution
$|y^{}_{\rm max}|$ for the massive gauge-bosons
for two generic  processes, $p\bar p \rightarrow W^+W^-W^\pm$
and $p \bar p \rightarrow W^+W^-Z$ in Fig.~\ref{FIG:YW}.
We see that the gauge-bosons
are produced in the central region, with  rapidities typically of $|y| \leq
1.5$.
The charged leptons from the $V$ decays then should be
mostly  within  a rapidity range less than 2.5.  Also, the transverse
momenta for the final state charged leptons will roughly peak around
$M_V^{}/2$ as the Jacobian peak for the two-body $V$-decay. Therefore,
we would not expect  to lose much rate after imposing realistic leptonic
cuts.

\section{Summary}

The triple gauge-boson production  processes must be quantified
for future experiments at an upgraded Tevatron in order
to explore physics beyond the SM, such as
studying quartic gauge-boson self-interactions, and
searching for new particles like gauginos, gluinos and squarks
in SUSY theories with leptonic final states.
We  have calculated the cross sections
for  the  triple gauge-boson production in the Standard Model (SM)
at Fermilab Tevatron energies.  At $\sqrt s=2$ TeV,
the cross sections are typically of  order 10 fb.
We  have also estimated the production
rates for  multi-lepton  final states from the gauge-boson decays.
The cross sections for
$p \bar p \rightarrow W^\pm \gamma\gamma, Z\gamma\gamma$
and $W^+W^- \gamma$ processes to pure leptonic final states
at $\sqrt s=2$ TeV are of order a few  fb,
resulting in a few dozen clean leptonic events
for an integrated luminosity of 10 fb$^{-1}$,
with the minimal cuts   of Eq.~(\ref{EQ:GCUT})
on the final state photons.
The pure leptonic modes from other gauge-boson channels
give significantly smaller rate. Especially, the trilepton modes
yield a cross section  of   order  0.1 fb at 2 TeV if there is no
significant Higgs boson contribution. For
$m_H^{} \simeq 2M_V^{}$, the $VVV$ production rates
could be enhanced by a factor of  $4-6$.

\acknowledgements
We would like to thank  David  Summers for discussions
on the $W\gamma\gamma$ process, and
Hiro Aihara  for a communication regarding detector issues.
This work was supported in part by
the U.S.\ Department of Energy under Contract No.
DE-FG03-91ER40647 and in part by a U.C.-Davis Faculty
Research Grant.
%
%

%
%
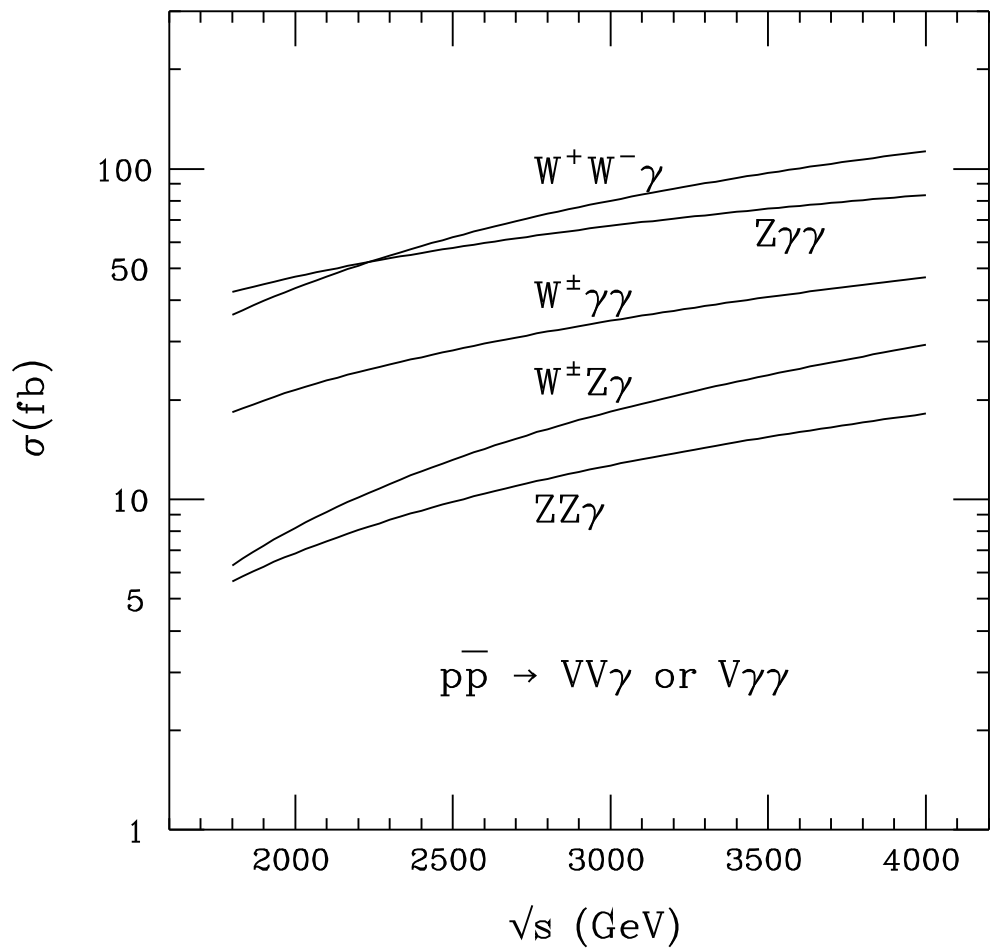
\begin{figure}
\caption{Total cross sections in units of fb versus the c.~m.~energy
$\protect \sqrt s$ for  $p \bar p \rightarrow V\gamma \gamma$
and  $p \bar p \rightarrow VV\gamma$ production (where $V=W^\pm,Z$).
Minimal cuts of Eq.~(\protect\ref{EQ:GCUT})
on the final state photons have been imposed.}
\label{FIG:VGG}
\end{figure}
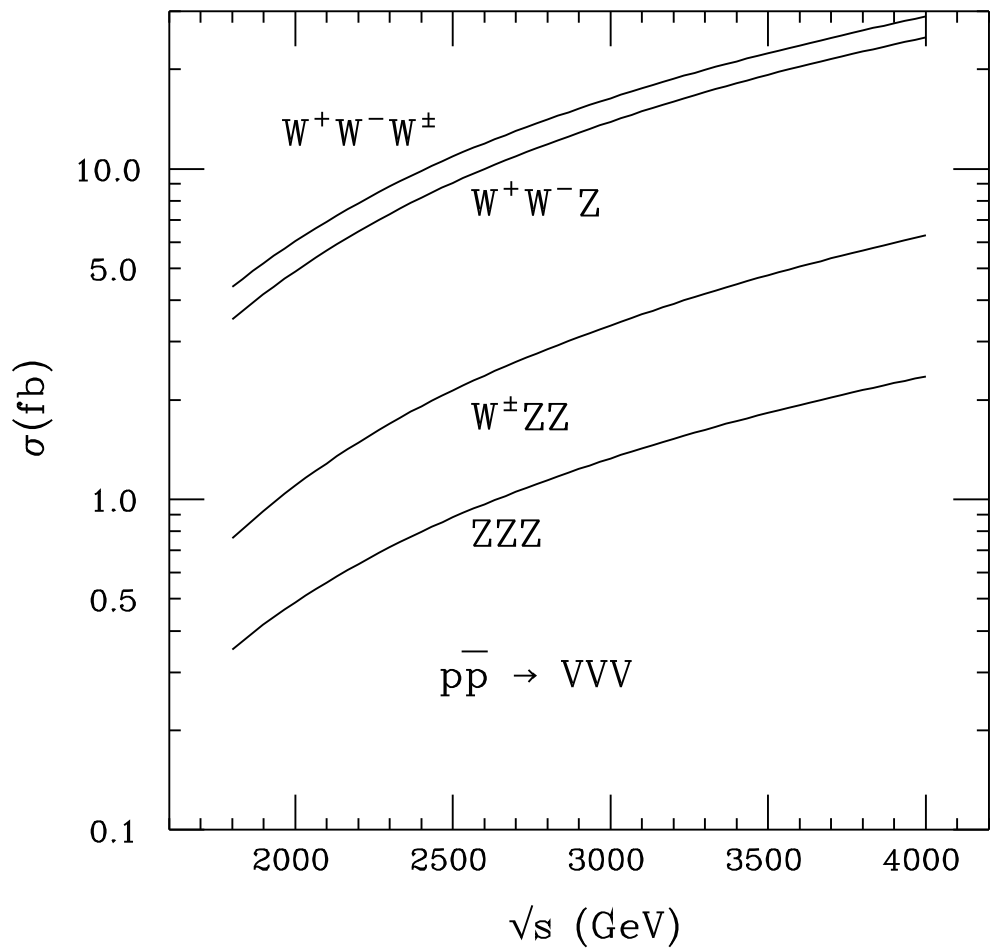
\begin{figure}
\caption{Total cross sections in units of fb versus the c.~m.~energy
$\protect \sqrt s$ for  $p \bar p \rightarrow VVV$ production.
Here $m_H^{}=100$ GeV is used.}
\label{FIG:VVV}
\end{figure}
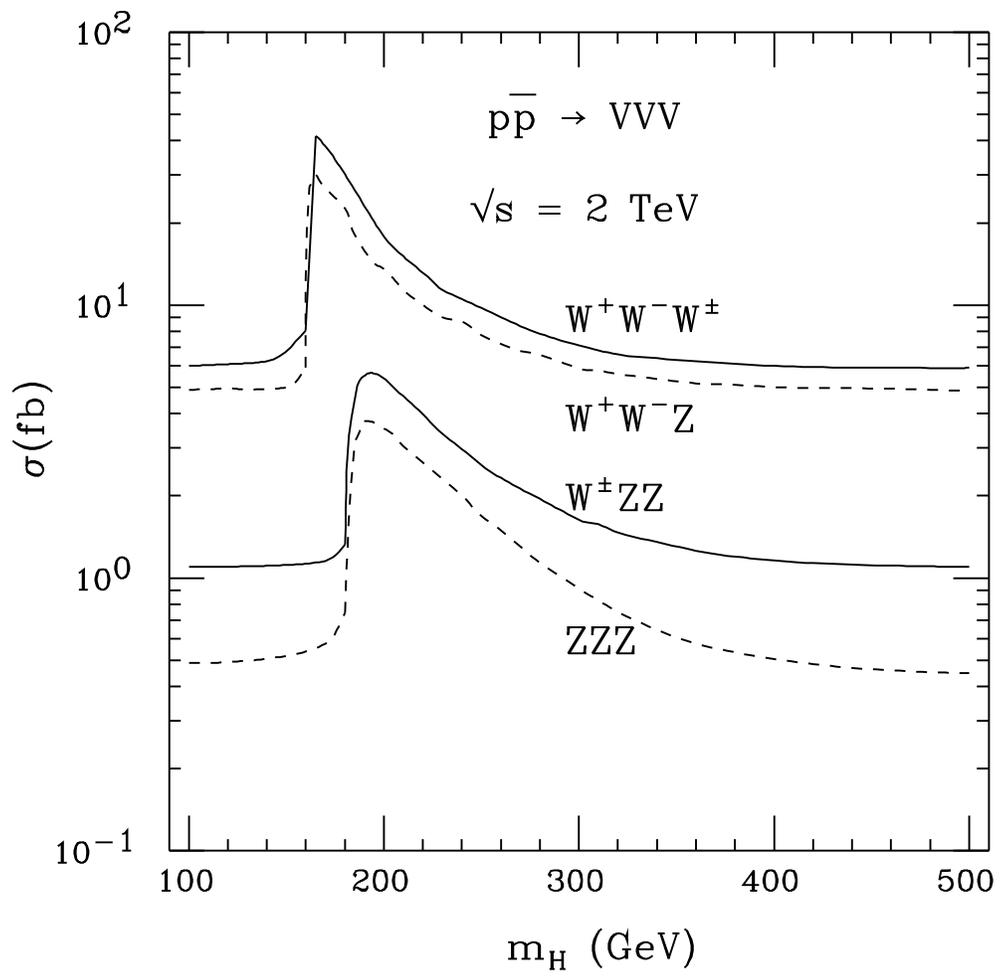
\begin{figure}
\caption{Total cross sections in units of fb versus the Higgs boson mass
$m_H^{}$ for  $p \bar p \rightarrow VVV$ production
at $\protect\sqrt s=2$ TeV. }
\label{FIG:HIGGS}
\end{figure}
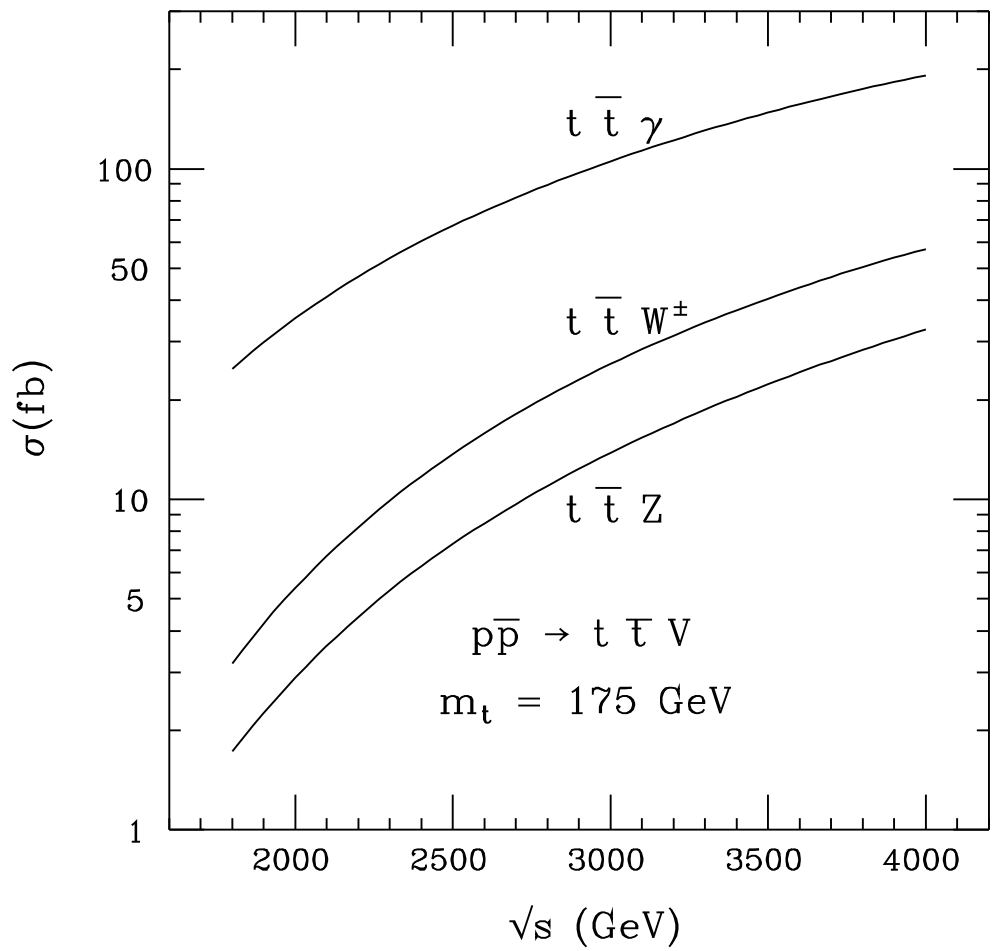
\begin{figure}
\caption{Total cross sections in units of fb versus the c.~m.~energy
$\protect \sqrt s$ for  $p \bar p \rightarrow t \bar t V$ production.
Here $m_t=175$ GeV is assumed.}
\label{FIG:TTV}
\end{figure}
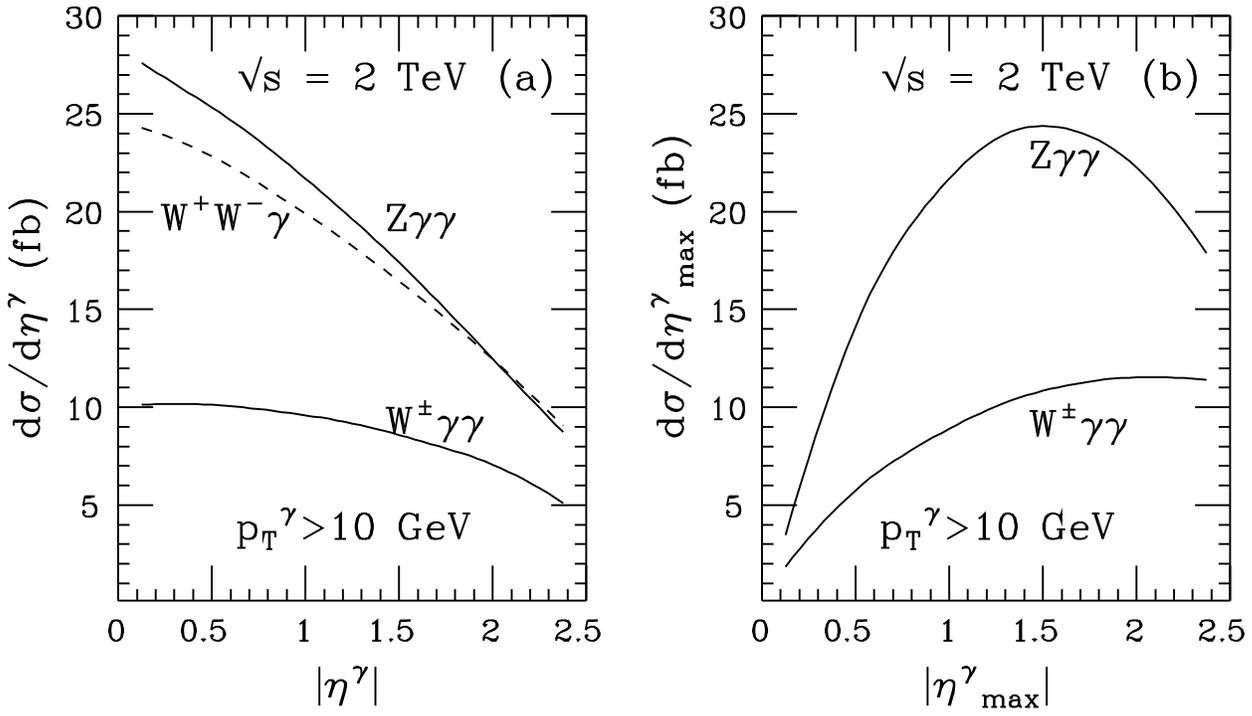
\begin{figure}
\caption{Photon pseudorapidity distribution at  $\protect \sqrt s$=2 TeV
for (a). $d\sigma/d\eta^\gamma$ and (b). $d\sigma/d\eta_{\rm max}^\gamma$
with $p_T^\gamma>10$ GeV.}
\label{FIG:ETA}
\end{figure}
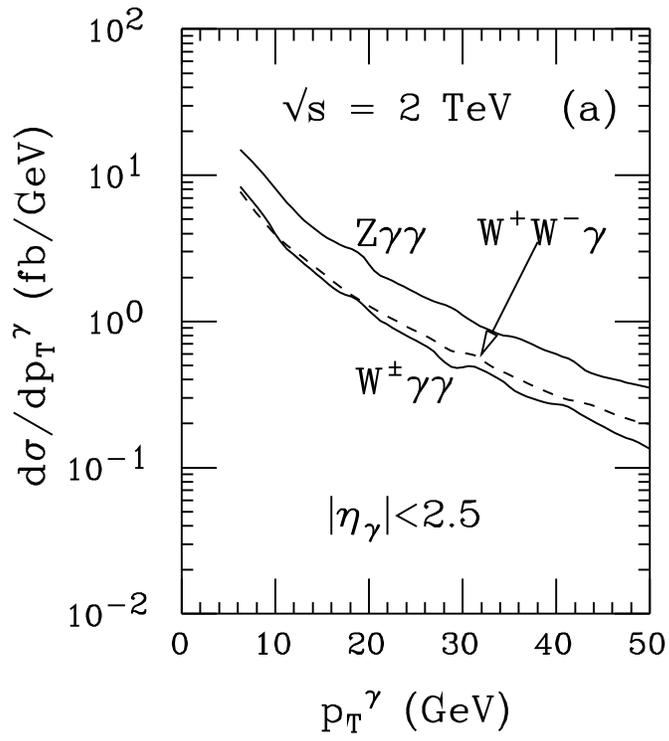
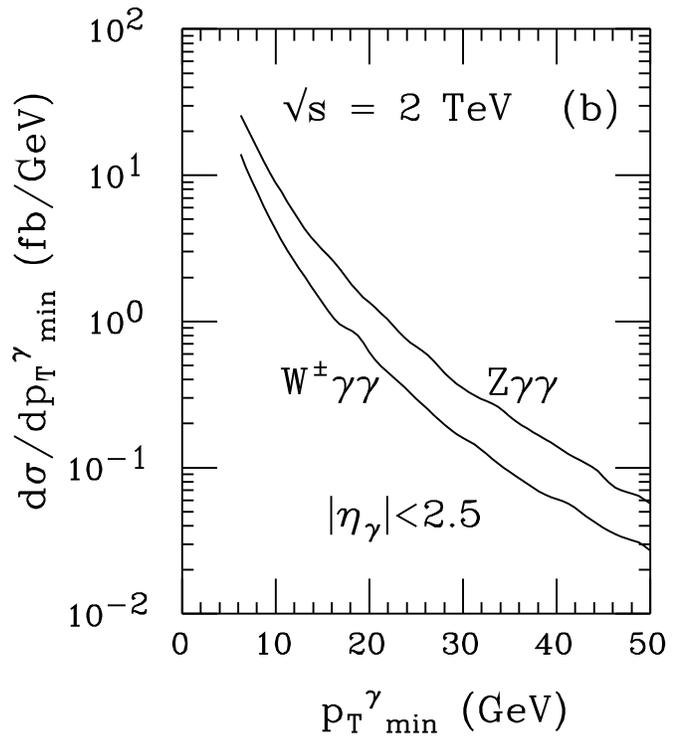
\begin{figure}
\caption{Photon transverse momentum distribution at  $\protect \sqrt s$=2 TeV
for (a). $d\sigma/dp_T^\gamma$ and (b). $d\sigma/dp_{T{\rm min}}^\gamma$
with $|\eta^\gamma|<2.5$.}
\label{FIG:PT}
\end{figure}
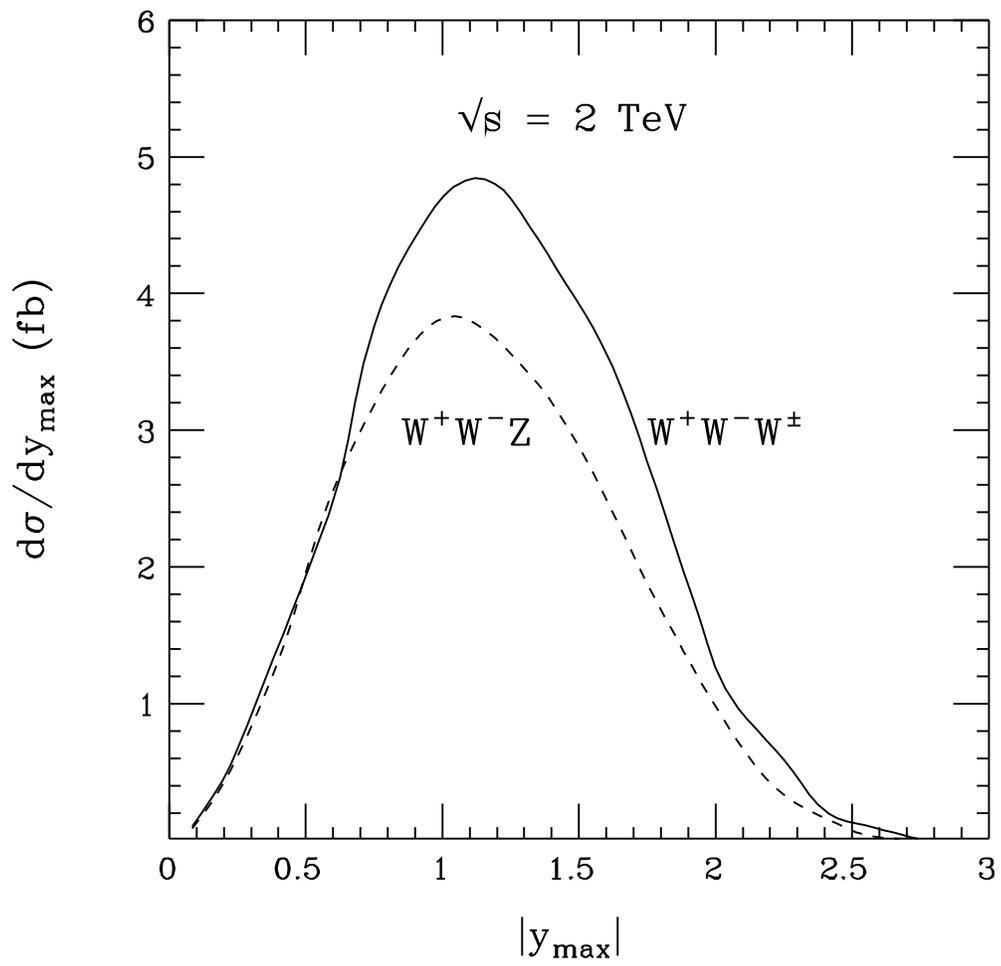
\begin{figure}
\caption{Maximum rapidity distribution  $d\sigma/dy^{}_{\rm max}$
for the massive gauge-bosons  at  $\protect \sqrt s$=2 TeV.
Here $m_H^{}=100$ GeV is used.}
\label{FIG:YW}
\end{figure}
\nopagebreak
%
%
\begin{table}[htb]
\centering
\caption[]{Cross sections in units of fb for triple gauge-boson production
in $p \bar p$ collisions for $\sqrt s$=2 TeV and $m_t=175$ GeV.
Minimal cuts of Eq.~(\ref{EQ:GCUT})
on final state photons have been imposed. The entries in the bottom row
include the pure leptonic branching fractions given in
Eq.~(\ref{EQ:BR}-\ref{EQ:BRVV}),
and $X$ denotes the missing neutrinos. }
\begin{tabular}{|c|c|c|c|c|}
 $W^\pm \gamma\gamma$ & $Z\gamma\gamma$ &
$W^+W^-\gamma$ \ ($t \bar t \gamma \rightarrow W^+W^-\gamma$)
& $W^\pm Z\gamma$  & $ZZ\gamma$  \\ \hline
$21 $ & $ 48 $ & $ 44 $ (36)&   $ 8.2 $ & $ 6.9 $ \\ \hline
 $l^\pm \gamma\gamma X$ & $l^+ l^-\gamma\gamma$ &
$l^+l^-\gamma X$ \ ($t \bar t \gamma \rightarrow l^+ l^-\gamma X$)
& $l^\pm l^+l^-\gamma X$  & $l^+l^-l^+l^-\gamma$  \\ \hline
$ 4.6 $ & $3.2 $ & $ 2.0 $ (1.7)&   $0.12 $ & $ 3.1\times 10^{-2} $ \\
\end{tabular}
\label{T:ONE}
\end{table}
\begin{table}[htb]
\centering
\caption[]{Cross sections in units of fb for triple gauge-boson production
in $p \bar p$ collisions for $\sqrt s$=2 TeV, $m_t=175$ GeV
and $m_H^{}=100$ GeV.
The entries in the bottom row  include the pure leptonic branching
fractions given in Eq.~(\ref{EQ:BRVVV}),
and $X$ denotes the missing neutrinos.}
\begin{tabular}{|c|c|c|c|}
$W^+ W^- W^\pm$  ($t \bar t  W^\pm \rightarrow W^+W^- W^\pm$)
&  $W^+W^- Z$   ($t \bar t  Z \rightarrow W^+W^- Z$)
&  $W^\pm ZZ$  & $ZZZ$  \\ \hline
$ 6.0 \ (5.4) $ & $4.9 \ (3.0) $ &   $ 1.1 $ & $ 0.49 $ \\ \hline
 $l^+ l^- l^\pm X$   ($t \bar t  W^\pm \rightarrow l^+ l^- l^\pm X$)
&  $l^+ l^- l^+l^-X$   ($t \bar t  Z \rightarrow l^+ l^- l^+l^-X$)
&  $l^\pm l^+l^-l^+l^-X$  & $l^+l^-l^+l^-l^+l^-$  \\ \hline
$ 5.9\times 10^{-2} \ (5.3\times 10^{-2}) $
& $1.5\times 10^{-2} \ (9.3\times 10^{-3}) $ &   $1.1\times 10^{-3} $
& $1.5\times 10^{-4} $ \\
\end{tabular}
\label{T:TWO}
\end{table}
%
%
\end{document}